\documentclass[12pt]{article}
\usepackage{epsf}
\def\beq{\begin{equation}}
\def\eeq{\end{equation}}
\def\ap#1#2#3 {Ann. Phys. (NY) {\bf#1} (19#2) #3}
\def\err#1#2#3 {{\it Erratum} {\bf#1} (19#2) #3}
\def\ib#1#2#3 {{\it ibid.} {\bf#1} (19#2) #3}
\def\ijmp#1#2#3 {Int. J. Mod. Phys. {\bf#1} (19#2) #3}
\def\jetp#1#2#3 {JETP Lett. {\bf#1} (19#2) #3}
\def\mpl#1#2#3 {Mod. Phys. Lett. {\bf#1} (19#2) #3}
\def\np#1#2#3 {Nucl. Phys. {\bf#1} (19#2) #3}
\def\pl#1#2#3 {Phys. Lett. {\bf#1} (19#2) #3}
\def\prep#1#2#3 {Phys. Rep. {\bf#1} (19#2) #3}
\def\prev#1#2#3 {Phys. Rev. {\bf#1} (19#2) #3}
\def\prl#1#2#3 {Phys. Rev. Lett. {\bf#1} (19#2) #3}
\def\sjnp#1#2#3 {Sov. J. Nucl. Phys. {\bf#1} (19#2) #3}
\def\spj#1#2#3 {Sov. Phys. JETP {\bf#1} (19#2) #3}
\def\spu#1#2#3 {Sov. Phys. Usp. {\bf#1} (19#2) #3}
\def\zp#1#2#3 {Zeit. Phys. {\bf#1} (19#2) #3}

\begin{document}
\begin{titlepage}
\begin{center}
{\Large \bf Theoretical Physics Institute \\
University of Minnesota \\}  \end{center}
\vspace{0.2in}
\begin{flushright}
TPI-MINN-99/07-T \\
UMN-TH-1742-99 \\
January 1999 \\
\end{flushright}
\vspace{0.3in}
\begin{center}
{\Large \bf  Relations between inclusive decay rates of heavy baryons
\\}
\vspace{0.2in}
{\bf M.B. Voloshin  \\ }
Theoretical Physics Institute, University of Minnesota, Minneapolis,
MN
55455 \\ and \\
Institute of Theoretical and Experimental Physics, Moscow, 117259
\\[0.2in]
\end{center}

\begin{abstract}

The dependence of inclusive weak decay rates of heavy hadrons on the
flavors of spectator light quarks is revisited with application to
decays of charmed and $b$ hyperons. It is pointed out that the
differences in the semileptonic decay rates, the differences in the
Cabibbo suppressed decay rates of the charmed hyperons, and  the
splitting of the total decay rates of the $b$ hyperons are all related
to the differences in the lifetimes of the charmed hyperons
independently of poor knowledge of hadronic matrix elements. The
approximations used in these relations are the applicability of the
expansion in the inverse of the charmed quark mass and the flavor SU(3)
symmetry.

\end{abstract}
\end{titlepage}

\section{Introduction}

The problem of unequal inclusive weak decay rates of hadrons containing
a heavy quark ($c$ or $b$) attracts considerable experimental and
theoretical interest ever since the first experimental indication
\cite{delco} of substantially different lifetimes of the charmed $D^0$
and $D^\pm$ mesons.  The total weak decay rates of charmed hadrons are
presently known to be vastly different, with the ratio of the longest
known lifetime to the shortest: $\tau(D^\pm) / \tau(\Omega_c) \sim 20$,
while the differences among the $b$ hadrons are much smaller. The
relation between the magnitude of the lifetime differences in the
charmed hadrons versus that in the $b$ hadrons reflects the fact that
these effects, associated with the light quark/gluon ``environment" of
the heavy quark $Q$ in a hadron, vanish as an inverse power of the heavy
mass $m_Q$, so that in the limit $m_Q \to \infty$ the `parton' picture
sets in, where the inclusive decay rate of a heavy hadron is given by
that of an isolated heavy quark.

Thus the problem of the differences in the rates can be approached
theoretically in the limit of large $m_Q$ in terms of a systematic
expansion {\rm [2 -6]} in $m_Q^{-1}$ for the decay rates of the hadrons.
The leading term in this expansion is the `parton' decay rate
$\Gamma_{part} \propto m_Q^5$, which sets the overall scale for the
decay rates of hadrons containing the quark $Q$ and which does not
depend on the specifics of the spectator light quarks (in baryons) or
the antiquark (in mesons). The first non-perturbative term, suppressed
with respect to the leading one by $m_Q^{-2}$, arises due to the kinetic
energy of the heavy quark in a hadron (time dilation effect) and due to
the cromomagnetic interaction of the heavy quark \cite{buv}. This term
generally gives a splitting of decay rates between heavy mesons and
baryons, and between baryons of different spin structure, however it
generates no splitting depending on the flavors of the spectator quarks
or antiquarks (e.g. between different $D$ mesons, or within the triplet
of baryons $\Lambda_c, ~ \Xi_c^0$ and $\Xi_c^+$). The flavor dependence
arises in the next order, $m_Q^{-3}$ relative to $\Gamma_{part}$, and
can be interpreted as due to two mechanisms: the weak scattering (WS)
and the Pauli interference (PI). The weak scattering corresponds to a
cross-channel of the decay, generically $Q \to q_1 \, q_2 \, {\overline
q}_3$, where either the quark $q_3$ is a spectator in a baryon and can
undergo a weak scattering off the heavy quark: $q_3 \, Q \to q_1 \,
q_2$, or an antiquark in meson, say ${\overline q}_1$, weak-scatters
(annihilates) in the process ${\overline q}_1 \, Q \to q_2 \, {\overline
q}_3$. The Pauli interference effect arises when one of the final
(anti)quarks in the decay of $Q$ is identical to the spectator
(anti)quark in the hadron, so that an interference of identical
particles should be taken into account. The latter interference can be
either constructive or destructive, depending on the relative spin-color
arrangement of the (anti)quark produced in the decay and of the
spectator one, thus the sign of the PI effect is found only as a result
of specific dynamical calculation.

Although the WS and PI effects carry the relative suppression by
$m_Q^{-3}$, they are found to be greatly enhanced by a large numerical
factor, typically $16 \, \pi^2/3$, reflecting mainly the difference of
the numerical factors in the two-body versus the three-body final phase
space, which makes these effects overwhelmingly essential in the charmed
hadrons, while greatly reduced in the heavier $b$ hadrons, as is
confirmed by the experimental data. In effect, the contribution of the
$O(m_Q^{-2})$ terms is significantly smaller than that of the
$O(m_Q^{-3})$ terms in the charmed hadrons and is slightly smaller in
the $b$ hadrons \cite{buv}. In particular the $O(m_Q^{-2})$ terms split
the decay rate of the $\Lambda_b$ from that of the $B$ mesons by only
about 2\%,
which is by far insufficient to explain the current data \cite{pdg} on
the ratio of the lifetimes: $\tau(\Lambda_b)/\tau(B)=0.79 \pm 0.05$ (for
a recent theoretical discussion see e.g. Ref. \cite{bsu}). A
quantitative estimate of the effects of the $O(m_Q^{-3})$ terms runs
into a problem of evaluating matrix elements over the hadrons of
four-quark operators of the type $({\overline Q} \, \Gamma_1 \, Q)
({\overline q} \, \Gamma_2 \, q)$ with certain spin-color matrix
structures $\Gamma_1$ and $\Gamma_2$. Although simple estimates within a
non-relativistic picture of the light quarks in hadrons (where these
operators reduce to the density of the light quarks at the location of
the heavy quark) allowed to correctly predict \cite{vs, vs1} the
hierarchy of the lifetimes of the charmed hadrons, these estimates are
obviously very unreliable for a quantitative description of the effect.
Neither can this approach explain the ratio $\tau(\Lambda_b)/\tau(B)$ to
be less than approximately $0.9$. In view of this difficulty it is quite
worthwhile to have a better understanding of the spectator flavor
dependent differences of the rates in a possibly more model-independent
way.

The purpose of this paper is to point out relations between some
inclusive decay rates of the charmed and $b$ baryons in the $(\Lambda_Q,
\, \Xi_Q)$ triplets, which do not require explicit knowledge of the
matrix elements of the four-quark operators, and rely only on the
general expansion in $m_Q^{-1}$ for the rates and on the flavor SU(3)
symmetry. Certainly, the latter symmetry is known to be not very
precise, however arguably the uncertainties due to the SU(3) violation
are substantially less than those brought in by the current model
assumptions about the hadronic matrix elements. Thus it may be expected
that an experimental verification of these relations can serve as a test
of the whole method based on the operator product expansion for weak
decay rates. To an extent, such approach was pursued in the prediction
\cite{mv} of a significant enhancement of the total semileptonic decay
rates of the $\Xi_c$ baryons over the same rate for the $\Lambda_c$ and
even greater enhancement of this rate for $\Omega_c$. That analysis was
further extended \cite{cheng, gm} to include the enhancement of the
Cabibbo suppressed semileptonic decays of $\Lambda_c$. Although those
papers used model considerations for the matrix elements of the
four-quark operators, in fact one can obtain, as shown in the present
paper, quantitative results for the $(\Lambda_c, \, \Xi_c)$ triplet
without resorting to models of quark dynamics in the baryons. Namely, it
will be shown that the difference of the semileptonic rates within the
$(\Lambda_c, \, \Xi_c)$ baryon triplet, both the dominant and the
Cabibbo suppressed, as well as the difference of the non-leptonic
Cabibbo suppressed decay rates, can all be expressed in terms of the
total lifetime differences within the same triplet in a
model-independent fashion, modulo the assumption of the flavor SU(3)
symmetry. In addition the differences of the lifetimes within the
triplet of the $b$ baryons $(\Lambda_b, \, \Xi_b)$ are also expressed
through the same differences for the charmed baryons, with a possible
extra uncertainty due to the quark mass ratio $m_b^2/m_c^2$.

Using the currently available data on lifetimes of the charmed hyperons,
the discussed effects are estimated to be quite large. In particular,
the conclusion of the previous analyses is confirmed that the
semileptonic decay rates of the $\Xi_c$ baryons should exceed by a
factor 2 to 3 the same rate for the $\Lambda_c$ hyperon. It is also
found that the lifetime of the $\Xi_b^-$ baryon can be longer than that
of $\Lambda_b$ by about 14\%, which is a very large effect for $b$
hadrons.

\section{Effective Lagrangian for spectator-flavor dependent effects in
decay rates}

The systematic description of the leading as well as subleading effects
in the inclusive decay rates of heavy hadrons arises \cite{sv0, vs, vs1}
through application of the operator product expansion in powers of
$m_Q^{-1}$ to the `effective Lagrangian' $L_{eff}$ related to the
correlator of two weak-interaction terms $L_W$:
\beq
L_{eff}=2 \,{\rm Im} \, \left [ i \int d^4x \, e^{iqx} \, T \left \{
L_W(x),
L_W(0) \right \} \right ]~.
\label{leff}
\eeq
In terms of $L_{eff}$ the inclusive decay rate of a heavy hadron $H_Q$
is given by the mean value\footnote{The non-relativistic normalization
for the {\it heavy} quark states is used throughout this paper: $\langle
Q | Q^\dagger Q | Q \rangle =1$.}
\beq
\Gamma_H=\langle H_Q | \, L_{eff} \, | H_Q \rangle~.
\label{lgam}
\eeq
The leading term in $L_{eff}$ describes the `parton' decay rate. For
instance, the term  in the non-leptonic weak Lagrangian $\sqrt{2} \, G_F
\, V ({\overline q}_{1 L} \gamma_\mu \, Q_L)({\overline q}_{2 L}
\gamma_\mu \, q_{3 L})$ with $V$ being the appropriate combination of
the CKM mixing factors, generates through eq.(\ref{leff}) the leading
term in the effective Lagrangian
\beq
L^{(0)}_{eff, \, nl} = |V|^2 \, {G_F^2 \, m_Q^5 \over 64 \, \pi^3} \,
\eta_{nl} \, \left ( {\overline Q} Q \right )~,
\label{lef0}
\eeq
where $\eta_{nl}$ is the perturbative QCD radiative correction factor.
In the limit $m_Q \to \infty$ this expression reproduces the `parton'
inclusive non-leptonic decay rate associated with the underlying process
$Q \to q_1 \, q_2 \, {\overline q}_3$, due to the relation $\langle H_Q
| {\overline Q} Q | H_Q \rangle \approx \langle H_Q | Q^\dagger Q | H_Q
\rangle =1$ with the approximate equality being valid up to corrections
of order $m_Q^{-2}$. In order to reproduce the complete expression of
order $m_Q^{-2}$ one should also include the first non-trivial term of
the OPE for $L_{eff}$ containing the dimension 5 chromomagnetic operator
$( {\overline Q} \, \sigma_{\mu \nu} G_{\mu \nu} \, Q)$ with the gluonic
field tensor $G_{\mu \nu}$. However neither this operator nor the
leading term $L_{eff}^{(0)}$ involve light quark operators, thus at this
level no splitting arises between the decay rates within flavor SU(3)
multiplets.

The dependence on the flavor of spectator quarks arises in the next term
of OPE for $L_{eff}$: $L_{eff}^{(3)}$. For the dominant Cabibbo
unsuppressed non-leptonic decays of the charmed quark, generated by the
underlying process $c \to s \, u \, {\overline d}$, this term reads as
\cite{vs1}:
\begin{eqnarray}
\label{l3nl}
&&L_{eff, \, nl}^{(3, 0)}= c^4 \,{G_F^2 \, m_c^2 \over 4 \pi} \,
\left \{
C_1 \, (\overline c \Gamma_\mu c)(\overline d \Gamma_\mu d) + C_2  \,
(\overline c \Gamma_\mu d) (\overline d \Gamma_\mu c) +\right .
\nonumber \\
&& C_3 \, (\overline  c \Gamma_\mu c +
{2 \over 3}\overline c \gamma_\mu \gamma_5 c) (\overline s \Gamma_\mu
s)+ C_4 \, (\overline  c_i \Gamma_\mu c_k +
{2 \over 3}\overline c_i \gamma_\mu \gamma_5 c_k)
(\overline s_k \Gamma_\mu s_i) +
\\ \nonumber
&& C_5 \, (\overline  c \Gamma_\mu c +
{2 \over 3}\overline c \gamma_\mu \gamma_5 c) (\overline u \Gamma_\mu
u)+ C_6 \, (\overline  c_i \Gamma_\mu c_k +
{2 \over 3}\overline c_i \gamma_\mu \gamma_5 c_k)
(\overline u_k \Gamma_\mu u_i)+ \\ \nonumber
&&\left. {1 \over 3} \,
\kappa^{1/2} \, (\kappa^{-2/9}-1) \, \left [ 2 \, (C_+^2 - C_-^2) \,
 (\overline c \Gamma_\mu t^a c) \,
j_\mu^a - (5C_+^2+C_-^2)
(\overline  c \Gamma_\mu t^a c +
{2 \over 3}\overline c \gamma_\mu \gamma_5 t^a c) j_\mu^a \right ]
\right \}~
\end{eqnarray}
with the coefficients $C_A, ~A=1, \ldots, 6$ given by
\begin{eqnarray}
&&C_1= C_+^2+C_-^2 + {1 \over 3} (1 - \kappa^{1/2}) (C_+^2-C_-^2)~,
\nonumber \\
&&C_2= \kappa^{1/2} \, (C_+^2-C_-^2)~, \nonumber \\
&&C_3=- {1 \over 4} \, \left [ (C_+-C_-)^2 + {1 \over 3}
(1-\kappa^{1/2})
(5C_+^2+C_-^2+6C_+C_-) \right] ~, \nonumber \\
&&C_4=-{1 \over 4} \, \kappa^{1/2} \, (5C_+^2+C_-^2+6C_+C_-)~, \nonumber
\\
&&C_5=-{1 \over 4} \, \left [ (C_++C_-)^2 + {1 \over 3} (1-\kappa^{1/2})
(5C_+^2+C_-^2-6C_+C_-) \right]~, \nonumber \\
&&C_6=-{1 \over 4} \, \kappa^{1/2} \, (5C_+^2+C_-^2-6C_+C_-)~.
\label{coefs}
\end{eqnarray}
In equations (\ref{l3nl}) and (\ref{coefs}) the following notation is
used: $\Gamma_\mu=\gamma_\mu \, (1-\gamma_5)$, $C_+$ and $C_-$ are the
standard coefficients in the QCD renormalization of the non-leptonic
weak interaction from $m_W$ down to the charmed quark mass:
$C_-=C_+^{-2}=(\alpha_s(m_c)/\alpha_s(m_W))^{4/b}$, where $b$, the
coefficient in the one-loop beta function in QCD, can be taken as
$b=25/3$ for the case of the charmed quark decay (see e.g. in the
textbook \cite{lbo}). Furthermore, the powers of the parameter $\kappa=
(\alpha_s(\mu)/\alpha_s(m_c))$ describe the so called `hybrid'
\cite{vs1, vs2} QCD renormalization of the operators from the
normalization scale $m_c$ down to a low scale $\mu$, and
$j_\mu^a=\overline u \gamma_\mu t^a u + \overline d \gamma_\mu t^a d +
\overline s \gamma_\mu t^a s$ is the color current of the light quarks
with $t^a = \lambda^a /2$ being the generators of the color SU(3).
Finally, the elements $V_{cs}$ and $V_{ud}$ of the CKM weak mixing
matrix are approximated here by the cosine of the Cabibbo angle, $c
\equiv \cos \theta_c$, hence the overall coefficient in eq.(\ref{l3nl})
is written as $c^4$.

The terms in eq.(\ref{l3nl}) with the flavor singlet operator $j_\mu^a$
produce no splitting of the decay rates within the flavor SU(3)
multiplets and thus are not of immediate relevance to the present
discussion. The rest of the terms containing operators with the $u$, $d$
and $s$ quarks with different coefficients are responsible for those
splittings. In terms of the physical interpretation of the effect the
operators with the $d$ quark describe the WS in charmed
hyperons\footnote{In mesons the same term describes the Pauli
interference of the $\overline d$ quark in the decays of $D^+$, which is
considered to be the dominant reason for the observed suppression of the
$D^+$ total decay rate.}, while the terms with the $u$ and $s$ quarks
correspond to the PI effects.

In order to discuss further in the paper the spectator effects on the
Cabibbo suppressed non-leptonic decays as well as on the semileptonic
ones, we will also need the expressions for the corresponding parts of
the effective Lagrangian $L_{eff}^{(3)}$. For the non-leptonic decays we
limit the present discussion to those suppressed by only one extra
factor of $s \equiv \sin \theta_c$, i.e. those generated by the quark
processes $c \to s \, u \, {\overline s}$ and $c \to d \, u \,
{\overline d}$. The Cabibbo-suppressed part of the non-leptonic
effective Lagrangian then can be found in the form:
\begin{eqnarray}
&&L_{eff, \, nl}^{(3, 1)}= c^2 \, s^2 \,{G_F^2 \, m_c^2 \over 4 \pi} \,
\left \{
C_1 \, (\overline c \Gamma_\mu c)(\overline q \Gamma_\mu q) + C_2  \,
(\overline c_i \Gamma_\mu c_k) (\overline q_k \Gamma_\mu q_i) +\right .
\nonumber \\
&& C_3 \, (\overline  c \Gamma_\mu c +
{2 \over 3}\overline c \gamma_\mu \gamma_5 c) (\overline q \Gamma_\mu
q)+ C_4 \, (\overline  c_i \Gamma_\mu c_k +
{2 \over 3}\overline c_i \gamma_\mu \gamma_5 c_k)
(\overline q_k \Gamma_\mu q_i) +  \\ \nonumber
&& 2 \, C_5 \, (\overline  c \Gamma_\mu c +
{2 \over 3}\overline c \gamma_\mu \gamma_5 c) (\overline u \Gamma_\mu
u)+ 2 \, C_6 \, (\overline  c_i \Gamma_\mu c_k +
{2 \over 3}\overline c_i \gamma_\mu \gamma_5 c_k)
(\overline u_k \Gamma_\mu u_i)+ \\ \nonumber
&&\left. {2 \over 3} \,
\kappa^{1/2} \, (\kappa^{-2/9}-1) \, \left [ 2 \, (C_+^2 - C_-^2) \,
 (\overline c \Gamma_\mu t^a c) \,
j_\mu^a - (5C_+^2+C_-^2)
(\overline  c \Gamma_\mu t^a c +
{2 \over 3}\overline c \gamma_\mu \gamma_5 t^a c) j_\mu^a \right ]
\right \}~
\label{l3nl1}
\end{eqnarray}
with the notation $(\overline q \, \Gamma \, q)= (\overline d \, \Gamma
\, d) + (\overline s \, \Gamma \, s)$.

The corresponding term in the effective Lagrangian for semileptonic
decays, generated by the quark-leton process $c \to s \, \ell^+ \, \nu$
and the Cabibbo-suppressed one: $c \to d \, \ell^+ \, \nu$ is given by
$\cite{mv, cheng, gm}$
\begin{eqnarray}
&&L_{eff, \, sl}^{(3)}= \nonumber \\
&&{G_F^2 \, m_c^2 \over 12 \pi} \, \left \{  c^2 \, \left [
L_1 \, (\overline  c \Gamma_\mu c +
{2 \over 3}\overline c \gamma_\mu \gamma_5 c) (\overline s \Gamma_\mu
s)+ L_2 \, (\overline  c_i \Gamma_\mu c_k +  {2 \over 3}\overline c_i
\gamma_\mu \gamma_5 c_k)
(\overline s_k \Gamma_\mu s_i) \right ] + \right. \nonumber \\
&& s^2 \, \left [
L_1\, (\overline  c \Gamma_\mu c +
{2 \over 3}\overline c \gamma_\mu \gamma_5 c) (\overline d \Gamma_\mu
d)+L_2 \, (\overline  c_i \Gamma_\mu c_k +
{2 \over 3}\overline c_i \gamma_\mu \gamma_5 c_k)
(\overline d_k \Gamma_\mu d_i) \right ] -  \nonumber \\
&& \left. \kappa^{1/2} \, (\kappa^{-2/9}-1) \,
(\overline  c \Gamma_\mu t^a c +
{2 \over 3}\overline c \gamma_\mu \gamma_5 t^a c) j_\mu^a
\right \} ~,
\label{l3sl}
\end{eqnarray}
where the coefficients $L_1$ and $L_2$ are
\beq
L_1=(\kappa^{1/2}-1), ~~~~~
L_2 = -   3\, \kappa^{1/2}~.
\label{coefl}
\eeq

In the next Section the general expressions in equations (\ref{l3nl}),
(\ref{l3nl1}) and (\ref{l3sl}) are used for an analysis of the relations
between the splittings of various inclusive decay rates within the
triplet of charmed baryons.

\section{Differences of inclusive decay rates for charmed baryons}

Estimates of the absolute effect of the flavor-dependent operators in
the effective Lagrangian $L_{eff}^{(3)}$ require evaluation of the
matrix elements of the four-quark operators over charmed hadrons. One
approach (for a review see e.g. Ref. \cite{bsu}) is to use a low
normalization point $\mu$ of the order of the confinement scale and
invoke a constituent quark model, simplifying it further to a
non-relativistic model, with possible additional (also non-relativistic)
input about the wave functions of the light quarks at the origin (see
e.g. Refs. \cite{grt, bs} and a rather general consideration in Refs.
\cite{nu, pu}). Needless to mention that such approach can
be used only for qualitative, or very approximate semi-quantitative
estimates, since it inevitably involves poorly controllable
approximations. In order to be able to find arguably more reliable
relations we do not attempt here an absolute evaluation of those matrix
elements, but rather use the flavor SU(3) properties of the operators in
$L_{eff}^{(3)}$ to relate the measurable splittings of the semileptonic
and the Cabibbo-suppressed inclusive decay rates in the charmed baryon
triplet to the splittins of the total decay rates. Namely, assuming the
flavor SU(3) symmetry, and the applicability of the heavy quark limit to
the charmed quark, it will be shown that the discussed splittings are
determined by only two independent matrix elements, which can be
expressed in terms of the differences in the measured total decay rates:
$\Delta_1 = \Gamma (\Xi_c^0) - \Gamma (\Lambda_c)$ and $\Delta_2= \Gamma
(\Lambda_c) - \Gamma (\Xi_c^+)$.

Proceeding with derivation of the relations we first notice that for
the triplet of the baryons $(\Lambda_c, \, \Xi_c)$ in the heavy quark
limit the spin of the charmed quark is not correlated with spinorial
characteristics of its light `environment'. Thus the operators with the
axial current of the $c$ quark give no contribution to the matrix
elements. The remaining flavor non-singlet structures in $L_{eff}^{(3)}$
involve only operators of the types $(\overline c \, \gamma_\mu \, c)
(\overline q \, \Gamma_\mu q)$ and $(\overline c_i \, \gamma_\mu \, c_k)
(\overline q_k \, \Gamma_\mu q_i)$ with $q$ being $d$, $s$ or $u$. Due
to the SU(3) symmetry the flavor non-singlet part of the matrix elements
of each of these types of operators in the baryon triplet is expressed
in terms of only one parameter. Indeed, the difference of the matrix
elements between the components of a $V$-spin doublet: $\Xi_c^0$ and
$\Lambda_c$, is contributed only by the $\Delta V=1$ combination of the
operators, proportional to $(\overline u \, \Gamma \, u) - (\overline s
\, \Gamma s)$, while the difference between the components of a $U$-spin
doublet: $\Lambda_c$ and $\Xi_c^+$, receives the same contribution only
from the $\Delta U =1$ operator $(\overline s \, \Gamma \, s) -
(\overline d \, \Gamma d)$.

Thus if one introduces two parameters $x$ and $y$ as
\begin{eqnarray}
x=\left \langle  {1 \over 2} \, (\overline c \, \gamma_\mu \, c) \left [
(\overline u \, \Gamma_\mu u) - (\overline s \, \Gamma_\mu s) \right]
\right \rangle_{\Xi_c^0-\Lambda_c} = \left \langle  {1 \over 2} \,
(\overline c \, \gamma_\mu \, c) \left [ (\overline s \, \Gamma_\mu s) -
(\overline d \, \Gamma_\mu d) \right] \right \rangle_{\Lambda_c -
\Xi_c^+}~,  \\ \nonumber
y=\left \langle  {1 \over 2} \, (\overline c_i \, \gamma_\mu \, c_k)
\left [ (\overline u_k \, \Gamma_\mu u_i) - (\overline s_k \, \Gamma_\mu
s_i) \right ] \right \rangle_{\Xi_c^0-\Lambda_c} = \left \langle  {1
\over 2} \, (\overline c_i \, \gamma_\mu \, c_k) \left [ (\overline s_k
\, \Gamma_\mu s_i) - (\overline d_k \, \Gamma_\mu d_i) \right] \right
\rangle_{\Lambda_c - \Xi_c^+}
\label{defxy}
\end{eqnarray}
with the shorthand notation for the differences of the matrix elements:
$\langle {\cal O} \rangle_{A-B}= \langle A | {\cal O} | A \rangle -
\langle B | {\cal O} | B \rangle$, the splitting of the inclusive decay
rates within the baryon triplet are expressed in terms of $x$ and $y$ as
follows. The differences of the dominant Cabibbo unsuppressed
non-leptonic decay rates are given by
\begin{eqnarray}
\delta_1^{nl, \,0} \equiv \Gamma^{nl}_{\Delta S =
-1}(\Xi_c^0)-\Gamma^{nl}_{\Delta S = -1}(\Lambda_c) = c^4 \, {G_F^2 \,
m_c^2 \over 4 \pi} \left [ (C_5 - C_3) \, x + (C_6 - C_4) \, y \right
]~, \nonumber \\
\delta_2^{nl, \,0} \equiv \Gamma^{nl}_{\Delta S =
-1}(\Lambda_c)-\Gamma^{nl}_{\Delta S = -1}(\Xi_c^+) = c^4 \, {G_F^2 \,
m_c^2 \over 4 \pi} \left [ (C_3 - C_1) \, x + (C_4 - C_2) \, y \right
]~.
\label{dnl0}
\end{eqnarray}
The once Cabibbo suppressed decay rates of $\Lambda_c$ and $\Xi_c^+$ are
equal, due to the $\Delta U =0$ property of the corresponding effective
Lagrangian $L_{eff, nl}^{(3,1)}$ (eq.(\ref{l3nl1})). Thus the only
difference for these decays in the baryon triplet is
\beq
\delta^{nl,1} \equiv \Gamma^{nl}_{\Delta S =0}
(\Xi_c^0)-\Gamma^{nl}_{\Delta S = 0 }(\Lambda_c) = c^2 \, s^2 \, {G_F^2
\, m_c^2 \over 4 \pi} \left [ (2\, C_5 - C_1 - C_3) \, x + (2 \, C_6 -
C_2 - C_4) \, y \right ]~.
\label{dnl1}
\eeq
The dominant semileptonic decay rates are equal among the two $\Xi_c$
baryons due to the isotopic spin property $\Delta I =0$ of the
corresponding interaction Lagrangian, thus there is only one non-trivial
splitting for these decays:
\beq
\delta^{sl,0} \equiv \Gamma^{sl}_{\Delta S = -1}(\Xi_c) -
\Gamma^{sl}_{\Delta S = -1} (\Lambda_c) = - c^2 {G_F^2 \, m_c^2 \over 12
\pi} \left [ L_1 \, x + L_2 \, y \right ]~.
\label{dsl0}
\eeq
Finally, the Cabibbo suppressed semileptonic decay rates are equal for
$\Lambda_c$ and $\Xi_c^0$, due to the $\Delta V =0$ property of the
corresponding interaction. Thus the only difference for these is
\beq
\delta^{sl,1} \equiv \Gamma^{sl}_{\Delta S = 0}(\Lambda_c) -
\Gamma^{sl}_{\Delta S = 0}(\Xi_c^+) = - s^2 {G_F^2 \, m_c^2 \over 12
\pi} \left [ L_1 \, x + L_2 \, y \right ]~.
\label{dsl1}
\eeq

Using the relations (\ref{dnl0}) - (\ref{dsl1}) one can express the
splittings of the total decay rates in terms of the two parameters $x$
and $y$ as
\begin{eqnarray}
\Delta_1= \delta_1^{nl, \,0} + \delta^{nl,1} + 2 \, \delta^{sl,0}~,
\nonumber \\
\Delta_2= \delta_2^{nl, \,0} - 2 \, \delta^{sl,0} + 2 \, \delta^{sl,1}~,
\label{dtot}
\end{eqnarray}
where the factor 2 for the semileptonic splittings takes into account
the decays with both $e \nu$ and $\mu \nu$ leptonic pairs, and the small
contribution of double Cabibbo suppressed non-leptonic decays is
neglected. Thus the unknown parameters $x$ and $y$ can be found in terms
of the measured differences $\Delta_1$ and $\Delta_2$ and used to
predict the splittings of inclusive decay rates for each experimentally
identifiable type of decays, described by the equations (\ref{dnl0}) -
(\ref{dsl1}).

It is quite satisfying to see that although the parameters $x$ and $y$
as well as the coefficients $C_A$ and $L_1$ and $L_2$ depend on the
arbitrarily chosen low normalization point $\mu$, in the resulting
relations between the physically measurable splittings the dependence on
$\mu$ cancels out, as it should be expected. In fact in the expression
for $\delta^{nl,1}$ in terms of the dominant non-leptonic splittings the
dependence on the QCD radiative effects cancels out altogether:
\beq
\delta^{nl,1}= {s^2 \over c^2} \, \left ( 2 \, \delta^{nl,0}_1 +
\delta^{nl,0}_2 \right )~,
\eeq
while the relation between the splitting of the dominant semileptonic
decay rates  and the dominant non-leptonic splittings, emerging after
excluding $x$ and $y$ in the equations (\ref{dsl0}) and (\ref{dnl0}),
reads as
\beq
\delta^{sl,0}= {1 \over c^2} \left [ {{5 \, C_-^2+ 5 \, C_+^2 - 2 \, C_+
\, C_-} \over 12 \, C_+ \, C_- \, (C_-^2 + 2 \, C_+^2)} \,
\delta_1^{nl,0} - {1 \over 3 \, (C_-^2 + 2 \, C_+^2)} \, \delta_2^{nl,0}
\right ]~.
\label{slnl}
\eeq
The coefficients in this relation depend only on the physical ratio of
the couplings $r=(\alpha_s(m_c)/\alpha_s(m_W))$. Moreover, this
dependence is in fact rather weak: in the absence of the QCD radiative
effects, i.e. with $r=1$, the coefficients in the square brackets in
eq.(\ref{slnl}) are $2/9 \approx 0.22$ and $-1/9 \approx -0.11$, while
with a realistic value $r \approx 2.5$ they are respectively 0.23 and
-0.09. Similar relative insensitivity of the numerical results to the
exact value of $r$ also holds for other relations between the observable
splittings. In subsequent numerical estimates the numerical values
$r=2.5$ and $s^2=0.05$ are used. We also use the data from Ref.
\cite{pdg} on the lifetimes of the charmed baryons in the form:
$\Gamma(\Lambda_c)=4.85 \pm 0.28 \, ps^{-1}$, $\Gamma(\Xi_c^+)=2.85 \pm
0.5 \, ps^{-1}$ and $\Gamma(\Xi_c^0)= 10.2 \pm 2 \, ps^{-1}$. (The error
bars on the lifetimes for the $\Xi_c$ hyperons are in fact not
symmetric. However the symmetry improves for the inverse quantities,
i.e. for
the total widths, and a close approximation to the errors in the widths
is used here in a symmetric fashion.)

Solving the equations (\ref{dtot}) for the Cabibbo dominant splitting of
the semileptonic widths yields
\beq
\delta^{sl, 0} = 0.13 \, \Delta_1 - 0.065 \, \Delta_2 \approx 0.59 \pm
0.32 \, ps^{-1}~,
\label{slres}
\eeq
and, obviously, for the Cabibbo suppressed semileptonic decays one has
$\delta^{sl, 1} = (s^2/c^2)\, \delta^{sl, 0}\approx 0.030 \pm 0.016 \,
ps^{-1}$. The solution of eqs.(\ref{dtot}) for the splitting of the
Cabibbo suppressed non-leptonic rates similarly gives
\beq
\delta^{nl, 1} = 0.082 \, \Delta_1 + 0.054 \, \Delta_2 \approx 0.55 \pm
0.22 \, ps^{-1}~.
\label{nl1res}
\eeq

\section{Splitting of lifetimes in the triplet of $b$ baryons}

Once the unknown baryonic matrix elements $x$ and $y$ are
phenomenologically determined through the differences of the total decay
rates in the charmed baryon triplet, one can use these parameters for
evaluating the differences of decay rates in the triplet of the $b$
baryons: $\Lambda_b, \, \Xi_b^0,$ and $\Xi_b^-$. Indeed, in the limit
where both the $c$ and the $b$ quarks are heavy the matrix elements of
the four-quark operators over the $b$ hyperons should be the same as for
the charmed ones, provided that the operators are normalized at a low
point $\mu$ which does not depend on the masses $m_c$ or $m_b$. For
proceeding in this manner we write here the expression \cite{vs1} for
the corresponding effective lagrangian for non-leptonic $b$ decays,
neglecting small kinematical effects $O(m_c^2/m_b^2)$ in the relevant
expressions for the two-body phase space of the pair $c \overline c$ or
$c q, \,$\footnote{The expression with these small terms included can be
found in Ref. \cite{ns}. Also only the CKM-dominant processes $b \to c
\, \overline u \, d$ and $b \to c \, \overline c \, s$ are taken into
account in order to keep the formulas simple. The contribution of
sub-dominant processes to the total rates is below the expected
uncertainty.}
\begin{eqnarray}
&&L_{eff, \, nl}^{(3, b)}= c^2 \, |V_{bc}|^2 \,{G_F^2 \, m_b^2 \over 4
\pi} \,
\left \{
{\tilde C}_1 \, (\overline b \Gamma_\mu b)(\overline u \Gamma_\mu u) +
{\tilde C}_2  \,
(\overline b \Gamma_\mu u) (\overline u \Gamma_\mu b) +\right .
\nonumber \\
&& {\tilde C}_5 \, (\overline  b \Gamma_\mu b +
{2 \over 3}\overline b \gamma_\mu \gamma_5 b) (\overline q \Gamma_\mu
q)+ {\tilde C}_6 \, (\overline  b_i \Gamma_\mu b_k +
{2 \over 3}\overline b_i \gamma_\mu \gamma_5 b_k)
(\overline q_k \Gamma_\mu q_i)+ \\ \nonumber
&&\left. {1 \over 3} \,
{\tilde \kappa}^{1/2} \, ({\tilde \kappa}^{-2/9}-1) \, \left [ 2 \,
({\tilde C}_+^2 - {\tilde C}_-^2) \,
 (\overline b \Gamma_\mu t^a b) \,
j_\mu^a - \right. \right. \\ \nonumber
&& \left . \left.
(5{\tilde C}_+^2+{\tilde C}_-^2 - 6 \, {\tilde C}_+ \, {\tilde C}_-)
(\overline  b \Gamma_\mu t^a b +
{2 \over 3}\overline b \gamma_\mu \gamma_5 t^a b) j_\mu^a \right ]
\right \}~,
\label{l3nlb}
\end{eqnarray}
where again the notation $(\overline q \, \Gamma \, q)= (\overline d \,
\Gamma \, d) + (\overline s \, \Gamma \, s)$ is used, and the
renormalization coefficients are determined by $\alpha_s(m_b)$ instead
of $\alpha_s(m_c)$:
$\tilde C_-=\tilde C_+^{-2}=(\alpha_s(m_b)/\alpha_s(m_W))^{4/b}$,
$\tilde \kappa=
(\alpha_s(\mu)/\alpha_s(m_b))$. The coefficients $\tilde C_A$ are
related to $\tilde C_-, ~\tilde C_+$, and $\tilde \kappa$ in the same
way as in eqs.(\ref{coefs}).

The dominant semileptonic decay $b \to c \, \ell \, \nu$ does not
involve light quarks, thus one expects no substantial splitting of the
semileptonic decay rates within flavor SU(3) multiplets of $b$ hadrons.
The non-leptonic effective lagrangian in eq.(\ref{l3nlb}) is symmetric
with respect to $s \leftrightarrow d$, i.e. it has $\Delta U =0$ (this
property is broken if the small kinematical effects of the $c$ quark
mass are kept in $L_{eff}^{(3,b)}$). Thus at this level there is no
splitting between the non-leptonic decay rates of $\Lambda_b$ and
$\Xi_b^0$. The splitting of the decay rate between either of these and
$\Xi_b^-$ is given in terms of $x$ and $y$ by
\beq
\Delta_b \equiv \Gamma(\Lambda_b)-\Gamma(\Xi_b^-)= c^2 \, |V_{bc}|^2
\,{G_F^2 \, m_b^2 \over 4 \pi} \, \left[ (\tilde C_5 - \tilde C_1) \, x
+ (\tilde C_6 - \tilde C_2) \, y \right]~.
\label{delb}
\eeq

One can notice that in the absence of any QCD radiative effects the
latter difference is simply related to $\delta_1^{nl,0}$ and
$\delta_2^{nl,0}$ for the charmed baryons:
\beq
\Delta_b = { |V_{bc}|^2 \over c^2} \, {m_b^2 \over m_c^2} \,
(\delta_1^{nl,0} + \delta_2^{nl,0})~.
\label{bc0}
\eeq
The QCD correction coefficients make the full expression somewhat more
lengthy:
\begin{eqnarray}
&&\Delta_b = { |V_{bc}|^2 \over c^2} \, {m_b^2 \over m_c^2} \, {1 \over
4 \, C_+ \, C_- \, (C_-^2+ 2 \, C_+^2) } \, \left \{ \left [ C_+^2 \,
\left ( (3  + 2 \, \xi ) \, \tilde C_-^2 + 4 \, \xi \, \tilde C_- \,
\tilde C_+ + 6 \, (1-  \xi) \, \tilde C_+^2 \right ) +  \right. \right.
\nonumber \\
&&\left. \left. 2 \,  C_+ \, C_- \left ( \tilde C_-^2 + 2 \tilde C_+^2
\right ) - C_-^2 \, \left ( (1-\xi) \, \tilde C_-^2 - 2 \, \xi \, \tilde
C_- \, \tilde C_+ + (2 +3 \, \xi) \, \tilde C_+^2 \right ) \right ] \,
\delta_1^{nl,0}+ \right. \nonumber \\
&& \left . 4\, C_+ \, C_- \, \left( \tilde C_-^2 + 2 \, \tilde C_+^2
\right ) \, \delta_2^{nl,0}
\right \}~,
\label{delbc}
\end{eqnarray}
where $\xi=(\tilde \kappa /\kappa)^{1/2} =
(\alpha_s(m_c)/\alpha_s(m_b))^{1/2}$.
One can again notice that the relation (\ref{delbc}) between physically
measurable quantities does not contain dependence on the low
normalization point $\mu$. Numerically, however the full expression is
not far from the simple approximation in eq.(\ref{bc0}): with a
realistic value $(\alpha_s(m_c)/\alpha_s(m_b)) \approx 1.25$ one finds
from eq.(\ref{delbc}):
$$
\Delta_b \approx { |V_{bc}|^2 \over c^2} \, {m_b^2 \over m_c^2} \, (0.91
\, \delta_1^{nl,0} + 0.93 \, \delta_2^{nl,0})~.
$$
When expressed in terms of the differences in the total decay rates
$\Delta_1$ and $\Delta_2$ for the charmed baryons, using
eq.(\ref{dtot}), the splitting of the decay rates within the $b$ baryon
triplet reads numerically as
\beq
\Delta_b \approx  |V_{bc}|^2  \, {m_b^2 \over m_c^2} \, (0.85 \,
\Delta_1 + 0.91 \, \Delta_2) \approx 0.015 \, \Delta_1 + 0.016 \,
\Delta_2 \approx 0.11 \pm 0.03 \, ps^{-1}~,
\label{dbres}
\eeq
which represents the estimate from the present analysis of the expected
suppression of the total decay rate of $\Xi_b^-$ with respect to that of
$\Lambda_b$, or $\Xi_b^0$.

\section{Discussion}

The relative differences of the lifetimes for charmed particles are
large, even within one flavor SU(3) triplet of the hyperons. Therefore
the assumption that these differences in the triplet are described by
just one term of the expansion in $m_Q^{-1}$ certainly requires
additional study. It should be noted however, that this assumption is
not necessarily flawed, since the discussed $O(m_Q^{-3})$ terms are
singled out by large numerical coefficient, and there is no reason for
recurrence of such anomaly further in the expansion. Thus the relations
between the splittings of the decay rates within the triplet charmed
hyperons $(\Lambda_c, \, \Xi_c)$ as well as of the $b$ hyperons
$(\Lambda_b, \, \Xi_b)$ may present a good testing point for
experimental study of this issue. As a consequence of large difference
in the lifetimes the additional effects, discussed here, are also
estimated to be quite large. The predicted difference in the
semileptonic decay rates between the $\Xi_c$ and $\Lambda_c$
(eq.(\ref{slres})) can be compared with the current data on the
semileptonic width of $\Lambda_c$:
$\Gamma_{sl} (\Lambda_c) = 0.22 \pm 0.08 \, ps^{-1}$. This comparison
confirms the conclusion of a previous analysis \cite{mv}, that the
semileptonic decay rates of the $\Xi_c$ hyperons can be larger than that
of $\Lambda_c$ by a factor of 2 or 3. Similarly, the small Cabibbo
suppressed semileptonic decay rate of $\Lambda_c$, should be enhanced by
the same factor \cite{gm} and may in fact constitute 10\% to 15\% of all
semileptonic decays of $\Lambda_c$. The effect in the Cabibbo suppressed
non-leptonic decays evaluated in eq.(\ref{nl1res}) can amount to more
than 10\% of the difference in the total non-leptonic decay rates of
$\Xi_c^+$ and $\Lambda_c$ and should be quite prominent, provided that
it would be possible to separate and measure the inclusive Cabibbo
suppressed rates experimentally.

Finally, the prediction of eq.(\ref{dbres}) for the difference of the
total decay rates of $\Lambda_b$ and $\Xi_b^-$ can be interesting in
relation to the mentioned earlier problem of the ratio $\tau
(\Lambda_b)/\tau (B)$. Indeed, the central number in eq.(\ref{dbres})
amounts to about 14\% of the total decay rate $\Gamma (\Lambda_b) = 0.81
\pm 0.05 \, ps^{-1}$, and a difference of such relative magnitude is
undoubtedly to be considered as very large for the $b$ hadrons.
If confirmed, this would indicate that the spectator effects in heavy
hyperons can be substantially larger, than usually expected, and may
shed some light on the problem of the $\Lambda_b$ versus $B$ lifetime.

\section{Acknowledgement}

This work is supported in part by DOE under the grant number
DE-FG02-94ER40823.

\end{document}